\documentstyle[sprocl,epsfig]{article}

\def\mathswitch#1{\relax\ifmmode#1\else$#1$\fi}
\def\mathswitchr#1{\relax\ifmmode{\mathrm{#1}}\else$\mathrm{#1}$\fi}

\newcommand{\PZ}{\mathswitchr Z}

\newcommand{\Pe}{\mathswitchr e}

\newcommand{\me}{\mathswitch {m_\Pe}}

\newcommand{\MZp}{\mathswitch {M_{\PZ'}}}
\newcommand{\GZp}{\mathswitch {\Gamma_{\PZ'}}}
\newcommand{\lesim}{\,\raisebox{-.1ex}{$_{\textstyle <}\atop^{\textstyle\sim}$}\,}

\begin{document}
\begin{flushright}
FERMILAB-Conf-04-150-T\\[1em]
\end{flushright}

\title{HOW TO STUDY WEAKLY COUPLED NEUTRAL VECTOR BOSONS%
\footnote{to appear in the proceedings of the International Conference on Linear
Colliders \mbox{(LCWS 04)}, Paris, France (April 19--23, 2004).}}

\author{AYRES FREITAS}

\address{Fermi National Accelerator Laboratory, Batavia, IL 60510-500, USA}


\maketitle\abstracts{
A weakly coupled new neutral gauge boson, forming a narrow resonance, can be
efficiently produced at $e^+e^-$ colliders through radiative return processes if
the collider energy is larger than the gauge boson mass. This contribution
analyzes the sensitivity of a future linear collider for weakly coupled gauge
bosons and briefly discusses how, in case of discovery, its properties can be
determined with high precision.
}

Many extensions of the Standard Model (SM) predict new neutral gauge bosons as part
of extended or additional gauge groups. Such a $Z'$ boson can have a mass as
low as the order of the $Z$ mass, in accordance with all experimental bounds,
if its couplings to the SM fermions are very weak \cite{leike:99}.
Current limits placed by searches at LEP  \cite{bogdan:03} and the Tevatron
\cite{TeVcur} could be improved by a future $e^+e^-$ high-energy linear
collider with high luminosity \cite{pap}. In this contribution the reach of a 1
TeV linear collider for new $Z'$ bosons is analyzed, focusing on very weakly
coupled $Z'$ bosons with masses below the center-of-mass energy.

It is assumed that mixing effects between the $Z$ and $Z'$ bosons are
negligible, so that no constraints on the $Z'$ boson arise from $Z$-pole data.
In this case, the most stringent bounds are obtained from direct $Z'$
production. For very small coupling strength, the $Z'$ will form a very narrow
resonance, that will lead to a significant signal in the process $e^+e^- \to f
\bar{f}$ only if the $Z'$ mass is close to the center-of-mass
energy, $\MZp \approx \sqrt{s}$. If instead $\MZp < \sqrt{s}$, the most
stringent constraints are obtained from the process involving the additional
radiation of an initial-state photon \cite{leike:99,bogdan:03,pap},
\begin{equation}
e^+e^- \to Z' + n\gamma \to f \bar{f} + n\gamma, \label{eq:radreturn}
\end{equation}
so that the $Z'$ boson can still be produced on-shell. In this analysis, the leading initial-state radiation effects due to large logarithms, $L = \log
s/\me^2$, have been included through the structure-function approach
\cite{structf} up to ${\cal O}(\alpha^2 L^2)$. Besides initial-state radiation,
beamstrahlung is taken into account, since it also leads to an effective reduction of
the invariant $Z'$ mass. The dominant background processes are
$
e^+e^- \to \gamma^*/Z^* + n\gamma \to f\bar{f} + n\gamma, \label{eq:bk}
$
and are included together with the
signal process (\ref{eq:radreturn}) into a Monte-Carlo simulation.
The detector response has been modeled by performing a simple binned analysis in
the $f \bar{f}$ invariant mass distribution, where the bin
size is determined by the momentum and/or energy resolution for the particular
particle species $f$. Due to the narrow width of the weakly coupled $Z'$ boson,
the signal will appear as an excess in a single bin.
For this study, the machine and detector resolution parameters have been taken
from the {\sc Tesla} design \cite{tesla}. The collider is likely to run at
different center-of-mass energies for various physics studies. Here the
following scenarios are taken into account: (i) $W$-boson threshold $\sqrt{s} =
170$ GeV, (ii) $t$-quark threshold $\sqrt{s} = 350$ GeV,
(iii) base-line high-energy design
$\sqrt{s} = 500$ GeV, and (iv) upgraded
high-energy version $\sqrt{s} = 1000$ GeV.

\begin{figure}[tb]
\epsfig{figure=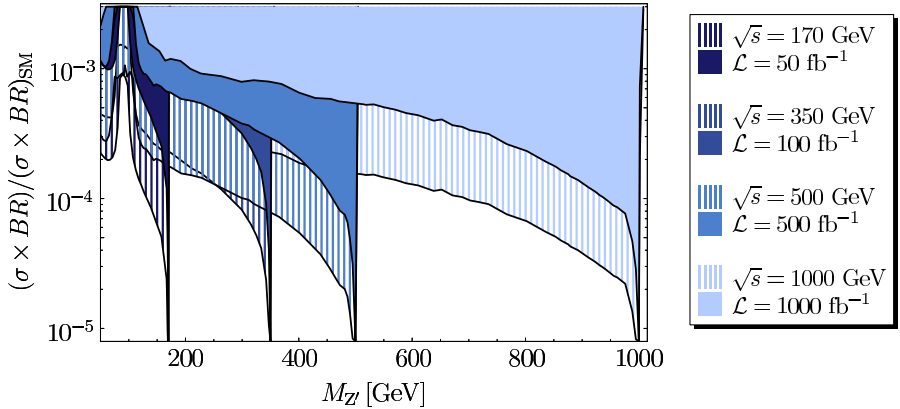, width=\textwidth, bb=75 460 515 660}
\vspace{-4ex}
\caption{Projected sensitivity of a $e^+e^-$ collider for exclusion at
95\% confidence level (hatched regions) and $5\sigma$ discovery (solid
regions) of a $Z'$ gauge boson in the di-jet channel. 
The reach in terms of production cross-section times
branching ratio, normalized to the value for a SM $Z$ boson, 
is shown as a function of the $Z'$ mass for various collider energies.}
\label{fg:bounds}
\end{figure}

Fig.~\ref{fg:bounds} shows as an example the projected
reach of a linear collider for the decay channel into light hadrons, $Z' \to q\bar{q},$ $q \neq t$.%
The plot shows that a $Z'$ boson with a signal rate about three orders of
magnitude smaller than for a gauge boson with SM $Z$ couplings can
be found throughout the range 50 GeV $< \MZp <$ 1 TeV, except near the $Z$
resonance. In particular classes of models, these limits can be translated into
limits on the $Z'$ couplings \cite{pap}. 

\begin{figure}[tb]
\epsfig{figure=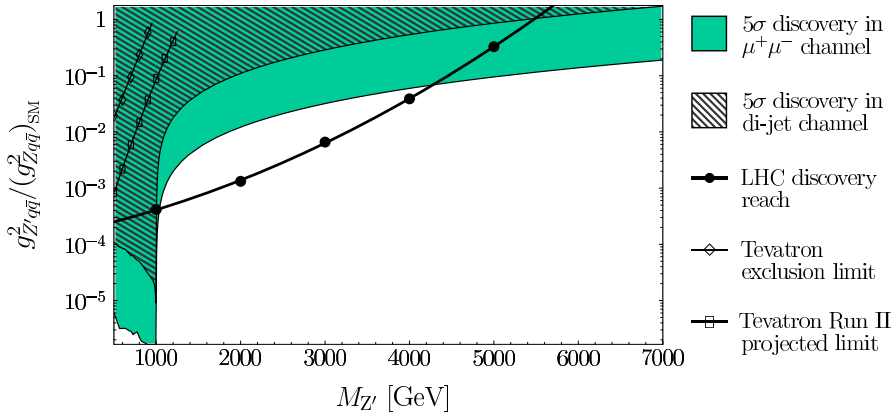, width=\textwidth}
\vspace{-4ex}
\caption{Projected discovery reach of a future $e^+e^-$ collider with $\sqrt{s}
= 1$ TeV and ${\cal L} = 1000 {\rm \ fb}^{-1}$ for a $Z_{\rm B-L}$ boson,
in comparison to the discovery reach of the
LHC, the present
limit from the Tevatron and
the expected exclusion reach at the end of Tevatron Run II.}
\label{fg:ZBL}
\end{figure}

For $Z'$ masses $\MZp > \sqrt{s}$, the  sensitivity in the process $e^+e^- \to
f\bar{f}$ falls off quickly, since the observable cross-section is modified
only through off-shell propagator effects from the $Z'$ resonance tail. For
sufficiently strong couplings, however, even $Z'$ bosons with masses
about an order of magnitude larger than $\sqrt{s}$ can be discovered and their
properties can be studied \cite{highmass}. Fig.~\ref{fg:ZBL} compares the
$5\sigma$ discovery limits for a $Z_{\rm B-L}$ boson at a 1 TeV linear
collider, inferred from total cross-section measurements, with the current and
expected coverage of the Tevatron Run II \cite{TeVcur,TeVfut} and the LHC \cite{lhc}.
The linear collider provides very good
sensitivity for $Z'$ bosons with masses below $\sqrt{s} = 1$ TeV and small
couplings, and for large values of $\MZp$, but relatively strong
couplings. In the intermediate region 1 TeV $< \MZp \lesim$ 4 TeV, the coverage
of the LHC is superior.

If a weakly coupled $Z'$ boson is discovered through the radiative-return
method, the properties of the new particle can be studied precisely by tuning
the collider energy to the $Z'$ resonance. As an example, a $Z_{\rm B-L}$ boson
with mass $\MZp = 400$ GeV and lepton coupling $\tilde{g}_l =
0.006$ is considered, leading to a signal cross-section in the $\mu^+\mu^-$
channel that is about 1000 times smaller than for $Z$ couplings. The total width
$\GZp \simeq 0.6$ MeV is too small to be directly resolved from the resonance
shape. However, from the analysis of decay ratios or
asymmetries one can determine couplings ratios with high precision. In the given
example, a measurement of the left-right and forward-backward asymmetries in the
channel $e^+e^- \to Z' \to \mu^+\mu^-$ allows to determine the ratios of left-
and right-handed $Z'ee$ and $Z'\mu\mu$ couplings with about 1\% accuracy, using
20 fb$^{-1}$ at $\sqrt{s} = 400$ GeV \cite{pap}.

The boson mass can be determined very precisely from a scan around the
resonance. Here the beam energy spread causes a strong correlation between the
mass $\MZp$ and the coupling $\tilde{g}_l$ of the $Z_{\rm B-L}$ boson. A fit of
these two parameters to a three-point scan at $\sqrt{s} = 399, 400, 401$ GeV,
spending 10 fb$^{-1}$ on each point, give the result  $\MZp =
400.0^{+0.007}_{-0.013}{^{+0.040}_{-0.040}}$. Here the first error is
statistical, while the second error is from systematic effects, dominated by
the uncertainty in the absolute beam energy.

It is also possible to search for invisible $Z'$ boson decays, which primarily
decay into neutrinos or other weakly interacting particles, using the process
$e^+e^- \to \gamma + {\rm missing\ energy}$, where a hard photon is required
for tagging \cite{invZ}. Since here the photon is not allowed to be in the
kinematical region collinear to the beam pipe, the expected sensitivity will be
roughly a factor $L = \log s/\me^2$ lower than the limits shown in
Fig.~\ref{fg:bounds}.

The measurement of the invisible $Z'$ decay modes allows in principle to
determine absolute branching ratios and couplings. However, the precision of
this method is limited due to the requirement of a non-collinear hard tagging
photon, so that  no collinear enhancement factor $L$ is present and the
collider energy cannot be tuned directly to the $Z'$ resonance. For very weakly
coupled $Z'$ bosons, as in the example presented above, it is therefore not
possible to obtain a significant signal for the invisible decays. Nevertheless,
for moderate coupling strength, for instance $\tilde{g}_l = 0.1$, a
determination of absolute branching ratios at the per-cent level is achievable
\cite{pap}.

\end{document}